\newcommand{\be}{\begin{equation}}
\newcommand{\ee}{\end{equation}}
\newcommand{\beeq}{\begin{eqnarray}}
\newcommand{\eeeq}{\end{eqnarray}}

\def\funp{{I\!\!P}}
\documentstyle[12pt,epsfig]{article}
\newlength{\dinwidth}
\newlength{\dinmargin}
\begin{document}
%\vskip 3cm
{\noindent
{July 1996} \hfill {INP Cracow  1734/PH}}
\vskip 3cm
\begin{center}
{\large \bf Subleading Reggeons in Deep Inelastic Diffractive Scattering} 
\vskip 0.1cm
{\large \bf at HERA}
\vskip 0.5cm
{K. Golec--Biernat and J. Kwieci\'nski } 
\vskip0.5cm
{\it 
H. Niewodnicza\'nski Institute of Nuclear Physics,}

{\it ul. Radzikowskiego 152, 31-342 Krak\'ow, Poland} 
\end{center}
\vskip1cm
\begin{abstract}
The contribution of subleading reggeons to the diffractive 
structure function 
${dF_{2}^D / dx_{\funp} dt}$ is estimated from
the soft physics data. This contribution leads in a natural way
to the violation of the factorization property of the diffractive
structure function. 

\end{abstract}
\vskip1cm
\setcounter{page} {1}
\thispagestyle{empty}
\newpage
The diffractive processes in deep inelastic
scattering, observed 
at the $ep$ collider HERA by the H1 and Zeus collaboration 
\cite{H1,ZEUS}, were interpreted in terms of the exchange of the leading
Regge trajectory corresponding to the soft pomeron 
 with a partonic substructure \cite{IS,CK,GS,GK}. In this interpretation the
diffractive interaction is treated as a two step process: an emission
of the soft pomeron from a proton 
 and a hard scattering of a virtual photon on a partonic
constituent of the pomeron. In this case the diffractive structure
function (DSF) is given in the factorizable form  
\be
\label{dsf}
{dF_{2}^D \over dx_{\funp} dt}(x,Q^2,x_{\funp},t)= 
         f^{\funp}(x_{\funp},t)~ F^{\funp}_{2}(\beta,Q^2)~.
\ee
Here $x$ is the Bjorken variable, $Q^2$ is the virtuality of the photon,
$x_{\funp}$ is the fraction of the momentum of the proton carried
away by the pomeron, $t$ is the pomeron virtuality and 
$\beta=x/x_{\funp}$. The function
 $f^{\funp}(x_{\funp},t)$ is the "pomeron flux"
describing the pomeron emission
and  $F^{\funp}_{2}(\beta,Q^2)$ is the pomeron structure
function.
 
The "pomeron flux" has the following form \cite{GK}
\be
\label{pomflux}
f^{\funp}(x_{\funp},t) = N~x_{\funp}^{1-2 \alpha_{\funp}(t)}~
                        {B_{\funp}^2(t) \over 16 \pi}~ ,
\ee
where $\alpha_{\funp}(t) = 1.1 + (0.25~GeV^{-2}) \cdot t$ 
is the pomeron trajectory
(with slightly increased value of the intercept 
$\alpha_{\funp}(0)$) and $B_{\funp}(t)$
is the pomeron coupling to a proton. The normalization factor $N$
was set to be equal to $2/\pi$.

The pomeron structure function  $F^{\funp}_{2}(\beta,Q^2)$
is related to the parton distributions in the pomeron 
in the same way as for the proton. 
The partonic structure of the pomeron was estimated in \cite{CK,GS,GK}
and independently fitted to
the diffractive HERA data \cite{H1,ZEUS} 
using the QCD evolution equations \cite{FITPARIS,GOJU,GO}.
Contrary to the proton case  a large gluonic component of the
pomeron was found at $\beta \rightarrow 1$.

In view of a new preliminary data, shown for the first time
by the H1 collaboration during the 1996 Eilat conference  
\cite{EILAT}, which suggest breaking of the factorization property
(\ref{dsf}) of the DSF, 
it is interesting to estimate the contribution of the subleading reggeons
to the factorization breaking of the DSF.
 This idea appeared for the first time among  other possibilities
in the H1 Collaboration talk in Eilat \cite{EILAT}.

We add to the DSF (\ref{dsf}) the "Regge" contribution
\be
\label{rsf}
{dF_{2}^R \over dx_{\funp} dt}(x,Q^2,x_{\funp},t)= 
         f^R(x_{\funp},t)~ F^R_{2}(\beta,Q^2)~,
\ee
where in this case $f^R(x_{\funp},t)$ is the "reggeon flux" and 
$F^R_{2}(\beta,Q^2)$ is the reggeon structure function. In principle
we should sum over the different Regge pole contribution and
include interference terms. As a first approximation, we assume
that the reggeon structure functions $F^R_{2}(\beta,Q^2)$ are
the same for all reggeons and we also neglect the interference terms
between different reggeons as well as 
between reggeons and the pomeron. In this case
\be
\label{fluxsum}
f^R(x_{I\!\!P},t) = \sum_{R_i} f^{R_i}(x_{I\!\!P},t)~,
\ee
and
\be
\label{rflux}
f^{R_i}(x_{I\!\!P},t) = N~x_{I\!\!P}^{1-2 \alpha_i(t)}~
                        {B_i^2(t) \over 16 \pi}~C_i(t)~ ,
\\ 
\ee
where $C_i(t)=4 cos^2(\pi \alpha_i(t)/2)$ or 
$C_i(t)=4 sin^2(\pi \alpha_i(t)/2)$ for even signature reggeons
$(f_2,a_2)$ or odd signature reggeons $(\rho,\omega)$
respectively. 
The functions $\alpha_i(t)$ are the reggeon trajectories, 
$B_i(t)$ denote the reggeon couplings to a proton and $N=2/\pi$. 
 
The dominant contribution to  sum (\ref{fluxsum}) comes
from the isoscalar exchanges of
$f_2$ and $\omega$ mesons
 which approximately lie on the same Regge trajectory
\be
\label{mtraj}
\alpha(t) \approx 0.5 + (1.0~GeV^{-2}) \cdot t~.
\ee
The corresponding couplings $B_i(0)$ can be deduced from the total
cross section data. The $f_2$ and  $\omega$ exchanges give the following
contribution to the total $pp$ and $p\bar p$ cross sections 
\beeq
\label{totcs}
\sigma_{pp}^{R} &=& sin(\pi \alpha(0))~(B_{f_2}^2 - B_{\omega}^2)~
             \Bigl ({s \over s_0}\Bigr)^{\alpha(0)-1}~,
\\
\sigma_{p\bar p}^{R} &=& sin(\pi \alpha(0))~ (B_{f_2}^2 + B_{\omega}^2)~
             \Bigl ({s \over s_0}\Bigr)^{\alpha(0)-1}~,
\eeeq
where $s_0=1 GeV^2$.
Using Donnachie and Landshoff parametrization \cite{DOLA1}, 
we obtain the following values of the couplings
\beeq
B^2_{f_2} &=& {\frac {98.39 + 56.08}{2}}~mb \approx 77.3~mb~,
\\
B^2_{\omega} &=& {\frac {98.39 - 56.08}{2}}~mb \approx 21.1~mb~.
\eeeq
We neglect the $t$ dependence of the couplings in our analysis.

The analytical form of the reggeon and pomeron 
structure function  $F^{R(\funp)}_{2}(\beta,Q^2)$ at small $\beta$
can be estimated from the triple Regge analysis of the diffractive
scattering, valid for a large mass $M_X$ of a diffractive system 
($\beta \rightarrow 0$). In this limit the pomeron structure function
is determined by the triple pomeron ($\funp\funp\funp$) coupling, while 
the reggeon-reggeon-pomeron ($RR\funp$) coupling determines the
reggeon structure function.
In both cases the 
structure functions have the same analytical form at small $\beta$
\vskip 0.1cm 
\be
\label{regsf}
F^{R(\funp)}_{2}(\beta,Q^2) = A_{R(\funp)}(Q^2)~\beta^{-0.08}~.
\ee

Due to the Regge factorization, the coefficient
$A_{R(\funp)}(Q^2)$ is a product of the $RR\funp$ ($\funp\funp\funp$)
coupling and the $Q^2$ dependent coupling of the pomeron to the 
virtual photons.
For the pomeron case 
the coefficient $A_{\funp}$ was estimated in \cite{GK} to be 
$A_{\funp}=0.03$  
at the scale $Q_0^2=4 GeV^2$. 
This relatively small value is a direct consequence of the small
magnitude of the triple pomeron coupling. 
The effective coupling controlling the
$RR\funp$ contribution to inclusive hadronic cross sections in
the triple Regge region is about one order of magnitude bigger then
that in the triple pomeron term \cite{FFOX}.
We do therefore expect that the 
$RR\funp$ coupling should also be
sigificantly bigger then the triple pomeron one. Thus, in our estimate
%we treat the $RR\funp$ coupling as a free parameter and 
we vary the coefficient
$A_R$ in formula (\ref{regsf}) within the limits $A_{R}=0.1-1.0$.
Finally, we extrapolated the parametrization of $F_2^{R}$
to the region of moderate and large values of $\beta$ by  multiplying the r.h.s of
formula (\ref{regsf}) by $(1-\beta)$.

We have also checked how the QCD evolution in $Q^2$ of the 
reggeon structure function $F_2^{R}$
influences the results, and found that
it   is not important, 
%for the binning in $(\beta, Q^2)$ given in Fig.1 
especially in view of the  above uncertainties.

In Fig.1 we show the DSF integrated over $t$, 
denoted by $F_2^{D(3)}(x_{\funp},\beta,Q^2)$,
as a function of $x_{\funp}$. The values of $\beta$ and $Q^2$
are those used in the H1 collaboration analysis
\cite{JUL}.
The solid lines correspond to the soft
pomeron contribution (\ref{dsf})
found in  analysis \cite{GO} of the published diffractive data
\cite{H1,ZEUS}. The dotted, dashed and 
dot-dashed lines correspond to the sum
of the pomeron and  reggeon contributions with the coefficient $A_{R}$
equal to $0.1, 0.5$ and $1.0$ respectively.

The DSF is now the sum of the two terms
\be
{dF_{2}^D \over dx_{\funp} dt}(x,Q^2,x_{\funp},t)= 
         f^{\funp}(x_{\funp},t)~ F^{\funp}_{2}(\beta,Q^2)
        + f^{R}(x_{\funp},t)~ F^{R}_{2}(\beta,Q^2)~,
\ee
where $f^{\funp}(x_{\funp},t) \not= f^{R}(x_{\funp},t)$. This fact leads
to a violation of the factoriztion property of the DSF, i.e. the DSF
is no longer a product of the $x_{\funp}$ dependent "flux factor"
and  $\beta$ dependent structure function.

To summarize, we estimated the subleading 
reggeon contribution to the DSF, 
based on the triple Regge limit of diffractive scattering. 
These corrections are important for small values of $\beta$ and for 
$x_{\funp}$ bigger then $10^{-2}$, however, the precise
value of the size of these corrections is difficult to estimate. The
subleading corrections  lead in a natural way to the violation
of the factorization property of the DSF

%%%%%%%%%%%%%%%%%%%%%%%%%%%%%%%%%%%%%%%%%%%%%%%%%%%%%%%%%%
%\newpage
\begin{figure}[htb]
   \vspace*{-1cm}
    \centerline{
     \psfig{figure=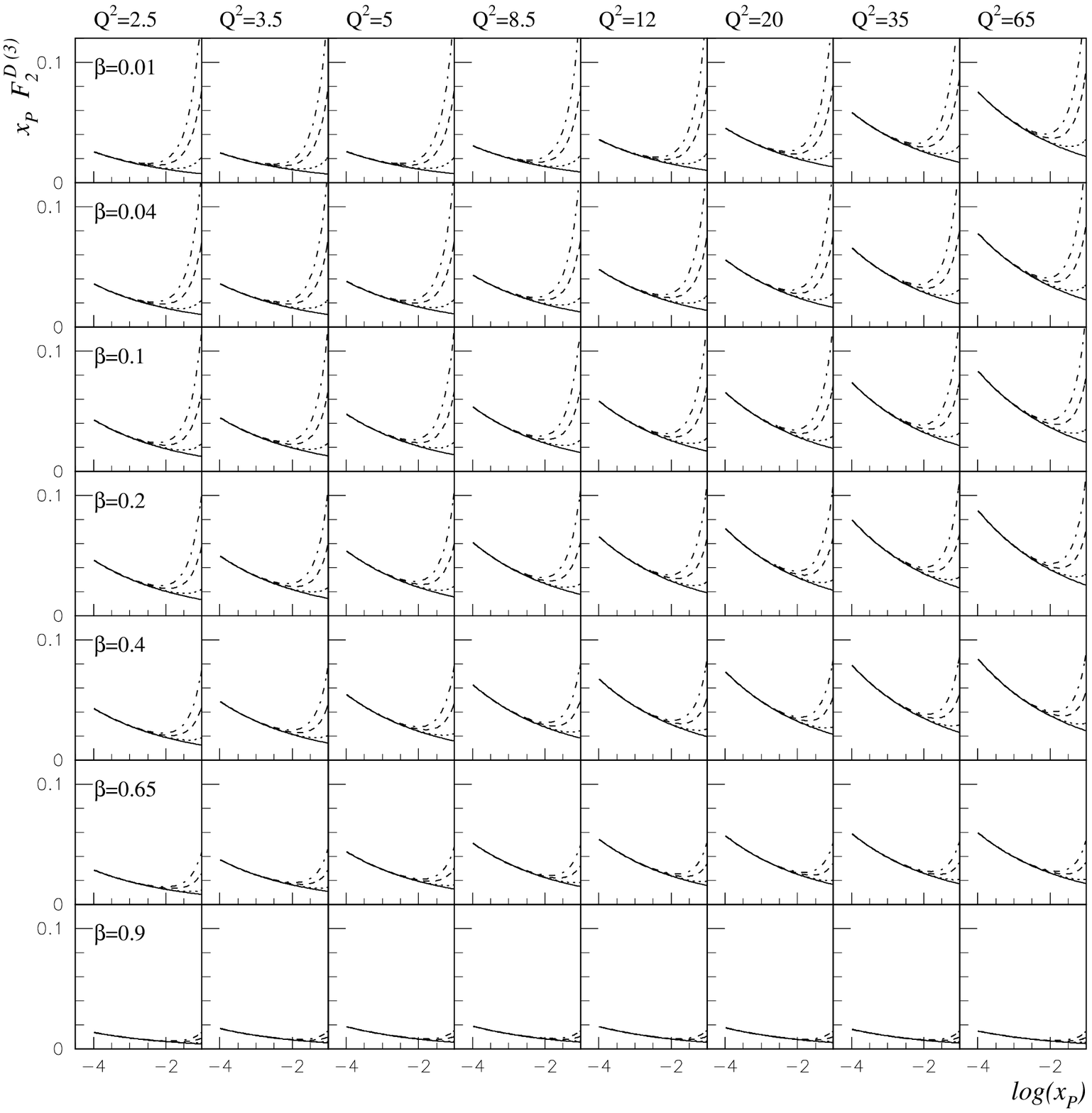,height=20cm,width=20cm}
               }
    \vspace*{-0.5cm}
     \caption{ 
$F_2^{D(3)}(x_{\funp},\beta,Q^2)$  as a function of $x_{\funp}$
for different $\beta$ and $Q^2$ values (given in $GeV^2$). The 
solid lines correspond to the soft pomeron contribution.  
The dotted, dashed and dot-dashed lines show the sum
of the pomeron and  reggeon contributions
 with the coefficient $A_{R}$
equal to $0.1, 0.5$ and $1.0$ respectively. 
}
\end{figure}

%%%%%%%%%%%%%%%%%%%%%%%%%%%%%%%%%%%%%%%%%%%%%%%%%%%%%

%\newpage
\vskip 1cm
{\bf Acknowledgments}
\par
We thank Albert de Roeck 
for a discussion which prompted this 
study and Julian P. Phillips for providing us details of the binning. 
 This research has been supported in part by the 
Polish State Committee for Scientific Research grants N0s 2 P03B 184 10,  
2 P03B 231 08, by Maria Sklodowska-Curie Fund II (No.PAA/NSF-94-158) 
and Stiftung f\"ur Deutsch-Polnische Zusammenarbeit, project no. 1522/94/LN.

\end{document}